\documentclass[a4paper]{jpconf}
\usepackage{graphicx,graphics,epsfig,hyperref}

\usepackage{color}
\usepackage{amsmath, mathbbol}
\usepackage{tikz,braids}
\begin{document}
\title{New type of solutions of Yang-Baxter equations,\, quantum entanglement and related physical models}

\author{Li-Wei Yu and Mo-Lin Ge}

\address{Theoretical Physics Division, S. S. Chern Institute of Mathematics, Nankai University, Tianjin, 300071. P.R.China}

\ead{yuliwei@mail.nankai.edu.cn; geml@nankai.edu.cn}

\begin{abstract}
Starting from the Kauffman-Lomonaco braiding matrix transforming the natural basis to Bell states, the spectral parameter describing the entanglement is introduced through Yang-Baxterization. It gives rise to a new type of solutions for Yang-Baxter equation, called the type-II that differs from the familiar solution called type-I of YBE associated with the usual chain models. The Majorana fermionic version of type-II yields the Kitaev Hamiltonian.  The introduced $\ell_1$ -norm leads to the maximum of the entanglement by taking the extreme value and shows that it is related to the Wigner's D-function. Based on the Yang-Baxter equation the 3-body S-Matrix for type-II is explicitly given. Different from the type-I solution, the type-II solution of YBE should be considered in describing quantum information. The idea is further extended to $\mathbb{Z}_3$ parafermion model based on  $SU(3)$ principal representation. The type-II is in difference from the familiar type-I in many respects. For example, the quantities corresponding to velocity in the chain models obey the Lorentzian additivity $\frac{u+v}{1+uv}$ rather than Galilean rule $(u+v)$. Most possibly, for the type-II solutions of YBE there may not exist RTT relation. Further more, for $\mathbb{Z}_3$ parafermion model we only need the rational Yang-Baxterization, which seems like trigonometric.  Similar discussions are also made in terms of generalized Yang-Baxter equation with  three spin spaces $\{1,\frac{1}{2},\frac{1}{2}\}$.
\end{abstract}

\section{Introduction}\label{sec1}

There is a long history for Yang-Baxter equation(YBE) based on exact solvable models, quantum inverse scattering methods, statistical models, quantum group, and related topics including S-matrix and CFT~\cite{yang1967some,yang1968matrix,baxter2007exactly,batchelor2007bethe,yang1991braid,takhtadzhan1979quantum,faddeev1981soviet,kulish1981lecture,kulish1982solutions,belavin1982solutions,drinfeld1986quantum,moore1989classical,jimbo1990yang,korepin1997quantum}.   YBE, together with the above theories, have been well established by 
many  inaugurators.

Typically, YBE takes the following form:
\begin{equation}\label{YBEMultiply}
\breve{R}_{12}(x) \breve{R}_{23}(xy) \breve{R}_{12}(y)=\breve{R}_{23}(y) \breve{R}_{12}(xy) \breve{R}_{23}(x),
\end{equation}
where $\breve{R}_{12}(x)=\breve{R}(x)\otimes I$,  $\breve{R}_{23}(x)=I\otimes \breve{R}(x)$ and $x$, $y$ stand for spectral parameters. For the familiar spin chain models usually $x=e^{iu}$, $y=e^{iv}$, then equation (~\ref{YBEMultiply}) becomes the following form:
\begin{equation}\label{YBEGailian}
\breve{R}_{12}(u) \breve{R}_{23}(u+v) \breve{R}_{12}(v)=\breve{R}_{23}(v) \breve{R}_{12}(u+v) \breve{R}_{23}(u),
\end{equation}
which means that the scattering obeys the Galilean additivity for velocities(rapidity) $u$ and $v$. 


 Originally, for the rational solution of Yang, the simplest  solution reads $(x=e^{iu})$
\begin{equation}
\breve{R}_{I}(u)=\mathbb{I}+uP,
\end{equation}
where $P$ represents the permutation operator and obeys $P^2=1$. To satisfy YBE, the parameters obey Galilean additivity $u+v$. We call this type of solutions the type-I solution. As usual in type-I solution, when $u\rightarrow\infty$, the YBE asymptotically reduces to braid relation. There are also trigonometric, chiral Potts model(including elleptic)~\cite{baxter1988new,au1987commuting,bazhanov1990chiral}, face model... in solving YBE.

On the other hand, in mathematical physics, the YBE is also associated with  braid group~\cite{frohlich1988statistics,yang1990braid}, 
which usually plays important roles in knot theory~\cite{kauffman1991knots}.  An N-strand braid group $\mathcal{B}_N$ has $N-1$ generators $\{B_i|i=1,2,...N-1\}$ with the following algebraic relations
\begin{eqnarray}
&&B_iB_j=B_jB_i \qquad \text{if} \,\, |i-j|\geq2,\\
&&B_iB_{i+1}B_i=B_{i+1}B_iB_{i+1}.\label{BRT}
\end{eqnarray}
The braid relation in equation (~\ref{BRT}) can be graphically expressed as follows, 
\begin{equation}
\begin{tikzpicture}[baseline=-\dimexpr\fontdimen22\textfont2\relax]
\draw(0,-1.1)node{\scriptsize $i$}(1,-1.1)node{\scriptsize $i+1$}(2,-1.1)node{\scriptsize $i+2$};
\braid[line width=1.3pt, height=15pt, number of strands=3, rotate=0,gray] (braid) at (0,1.3) s_1^{-1}s_2^{-1}s_1^{-1}; 
\end{tikzpicture}=
\begin{tikzpicture}[baseline=-\dimexpr\fontdimen22\textfont2\relax]
\draw(0,-1.1)node{\scriptsize $i$}(1,-1.1)node{\scriptsize $i+1$}(2,-1.1)node{\scriptsize $i+2$};
\braid[line width=1.3pt, height=15pt, number of strands=3, rotate=0,gray] (braid) at (0,1.3) s_2^{-1}s_1^{-1}s_2^{-1}; 
\end{tikzpicture}\, .
\end{equation}
Indeed, this graphical braid relation, is nothing but the Reidemeister move.


The local unitary representation of braid group  $\mathcal{B}_N$ in tensor product space $(\mathbb{C}^k)^{\otimes N}$ takes the following form,
\begin{equation}
 B_{i}=I^{\otimes i-1}\otimes B \otimes I^{\otimes N-i-1},\textrm{(Local representation)}
 \end{equation}
 where $B$ is a $k^2\times k^2$ braid matrix.

Parametrization(or Yang-Baxterization) of the braid matrix $B\rightarrow \breve{R}(x)$ yields the YBE~\cite{jimbo1985quantum,jones1989certain,ge1991explicit,cheng1991yang,li1993yang,ge1993trigonometric,ge1994yang}, 
\begin{equation}\label{YBEGailian}
\breve{R}_{12}(x) \breve{R}_{23}(xy) \breve{R}_{12}(y)=\breve{R}_{23}(y) \breve{R}_{12}(xy) \breve{R}_{23}(x).
\end{equation}
Specifically, in this paper we are interested in this braid matrix $B$, say $4\times4$~\cite{dye2003unitary}
\begin{equation}\label{type II}
   B = \tfrac{1}{\sqrt{2}}\left[
   \begin{array}{cccc}
   1 & 0 & \ 0 & \ 1 \\
   0 & 1 & \ 1 & \ 0 \\
   0 & -1 & \ 1 & \ 0 \\
   -1 & 0 & \ 0 & \ 1
   \end{array} \right],
\end{equation}
which is associated to extra special 2-group~\cite{franko2006extraspecial}.
In 2004, Kauffman and Lomonaco observed that applying the braid matrix on the 2-qubit natural tensor bases, it yield~\cite{Kauffman2004}
 \begin{equation}\label{braiding matrix}
  B\left[
  \begin{array}{c}
  |00\rangle \\|01\rangle \\|10\rangle \\|11\rangle
  \end{array}
  \right]=\tfrac{1}{\sqrt{2}}\left[
  \begin{array}{c}
  |00\rangle+|11\rangle \\|01\rangle+|10\rangle \\|10\rangle-|01\rangle \\|11\rangle-|00\rangle
  \end{array}
  \right].
\end{equation}
We see that the braid matrix  is related to the 2-qubit maximal entanglement. This braid matrix is also named Bell matrix due to its relation to the Bell basis.

Later we will show that the type of representation is closely related to the Ising anyon system  as well as Majorana operators~\cite{wang2010topological}. Based on the Majorana representation of the braid group, we extend the relationship between braid matrix and Bell bases, and obtain the general relationship between this braid representation and N-qubit quantum entanglement~\cite{yu2018local}.

This paper is organized as follows. In section~\ref{sec1},   we briefly introduce the background of YBE and the braid matrix in describing entanglement. In section~\ref{sec2}, we review the $4\times4$ type-II solution of YBE and its mathematical physical consequences. In section~\ref{sec3}, we extend to the $9\times9$ Yang-Baxter solution, and discuss the corresponding properties. In section~\ref{sec4}, we give a solution of non-standard YBE with three lattices spin $\{1,\frac{1}{2}, \frac{1}{2}\}$. At last section, we make conclusions and discussions about our results.

\section{$4\times4$ Type-II Solution of Yang-Baxter Equation}\label{sec2}

In the previous section, we have introduced the Bell braid matrix and the Yang-Baxterization approach.  In this section, we will introduce the $4\times4$ type-II solution of  YBE and some of  its properties. The $4\times4$ type-II solution comes from the Yang-Baxterization of Bell matrix $B$~\cite{chen2007braiding,chen2008berry}, 
 \begin{equation}
  \breve{R}_{12}(\theta_1) \breve{R}_{23}(\theta_2) \breve{R}_{12}(\theta_3)=\breve{R}_{23}(\theta_3) \breve{R}_{12}(\theta_2) \breve{R}_{23}(\theta_1),
 \end{equation}
 with the $\breve{R}$-matrix {(Type-II solution)}
 \begin{equation}
 \begin{aligned}\label{Type-II}
   \breve{R}(\theta,\varphi) &= \left[
   \begin{array}{cccc}
   \cos\theta & 0 & \ 0 & \ \sin\theta e^{-i\varphi} \\
   0 & \cos\theta & \ -\sin\theta & \ 0 \\
   0 & \sin\theta & \ \cos\theta & \ 0 \\
   -\sin\theta e^{i\varphi} & 0 & \ 0 & \ \cos\theta
   \end{array} \right],\\
   &=e^{-i\theta [\sigma^y\otimes \sigma^x]},\, (\textrm{for} \,\varphi=0),\\
   &=\cos\theta*(\mathbb{I}_4+\tan\theta \, M), \quad (M^2=-\mathbb{I}_4, \tan\theta=u)\\
   &=\frac{1}{\sqrt{1+u^2}}*(\mathbb{I}_4+u \, M).
      \end{aligned}
\end{equation} 
This is noting but the rational solution of YBE. Then we obtain the parameter constraint from type-II solution of YBE 
\begin{equation}\label{lorentzadd0}
   \tan\theta_2=\frac{\tan\theta_1+\tan\theta_3}{1+\tan\theta_1 \tan\theta_3}.
\end{equation}
What interesting here is the rational parameter $u=\tan\theta$, usually regarded as velocity or rapidity, obeys {Lorentzian additivity} instead of Galilean additivity usually appeared in the type-I solutions. However, the real physical meaning of $\theta$ remains to be discovered.
 
 Applying $\breve{R}(\theta)$ on the 2-qubit natural basis, it  yields~\cite{chen2007braiding}
\begin{equation}
  \breve{R}(\theta)\left[
  \begin{array}{c}
  |00\rangle \\|01\rangle \\|10\rangle \\|11\rangle
  \end{array}
  \right]=\left[
  \begin{array}{c}
  \cos\theta|00\rangle-\sin\theta |11\rangle \\\cos\theta|01\rangle-\sin\theta|10\rangle \\\sin\theta|01\rangle+\cos\theta|10\rangle \\ \cos\theta|11\rangle+\sin\theta |00\rangle
  \end{array}
  \right]\quad (\varphi=0). 
\end{equation}
The states on the right hand side are 2-qubit pure states with any entangled degree. Hence our $\breve{R}$-matrix describes the 2-qubit entanglement.  In this sense, the  physical meaning of  $\theta$ is 2-qubit continuous entangled degree.

When $\theta=\pi/4$,  $\breve{R}(\theta)$ turns back to the braid matrix $B$, which corresponds to the maximum of entanglement. And then the YBE turns out to be the braid relation.
 
 Based on the type-II solution, our group have obtained a series of  interesting mathematical physics results.  Now we introduce some properties associated to the $4\times4$ type-II solution. 
 
%

\subsection{ Yang-Baxter Hamiltonian and Berry phase}
The Yang-Baxter solution $\breve{R}$ we have obtained is unitary. Hence, one can reasonably  regard it as the system evolution operator along time.  Taking the Schr\"{o}dinger equation $i\hbar\tfrac{\partial}{\partial t}|\psi(t)\rangle=\hat{H}(t)|\psi(t)\rangle$ into account, one obtains:
\begin{equation}
i\hbar\tfrac{\partial}{\partial t}[\breve{R}_i|\psi(0)\rangle]=\hat{H}(t)\breve{R}_i|\psi(0)\rangle.
\end{equation}
If we suppose only $\varphi$ in $\breve{R}(\theta,\varphi)$ time-dependent, then the 2-body Hamiltonian is given as follows~\cite{chen2007braiding}
\begin{equation}\label{2BodyH}\begin{aligned}
\hat{H}_{i}^{(\varphi)}=&-i\hbar\frac{\partial \breve{R}_i(\theta,\varphi(t))}{\partial t}\breve{R}_i^\dag(\theta,\varphi(t)),\\
=&\hbar\dot{\varphi}\sin\theta \left[\frac{\sin\theta}{2}(\sigma^{z}_i+\sigma^{z}_{i+1})-\cos\theta (e^{-i\varphi}\sigma^+_i\sigma^+_{i+1}+e^{i\varphi}\sigma^-_i\sigma^-_{i+1})\right].
\end{aligned}\end{equation}
It describes the two spin-$\frac{1}{2}$ interaction. If we extend the 2-body Hamiltonian into the chain model, it is obviously superconductive under Jordan-Wigner transformation. This Hamiltonian has been well studied in some respects, such as in describing entanglement and Berry phase~\cite{chen2007braiding}, in quantifying quantum phase transition by ground state entanglement~\cite{hu2008exact}, 3-body interactions~\cite{yu2014factorized}, and so on.
 
 Among all the physical consequences of Yang-Baxter Hamiltonian, one interesting result is about the relationship between  Berry phase and entanglement. 
   
Under the adiabatic evolution of the system described in equation (~\ref{2BodyH}), we have totally four eigenstates, with two eigenstates zero-energy. Another two eigenstates with  non-zero energy can be written as
\begin{equation}
\begin{aligned}
&|\psi_+\rangle=\cos(\frac{\theta}{2}-\frac{\pi}{4})e^{-i\varphi}|\uparrow\uparrow\rangle-\sin(\frac{\theta}{2}-\frac{\pi}{4})|\downarrow\downarrow\rangle,\\
&|\psi_-\rangle=\sin(\frac{\theta}{2}-\frac{\pi}{4})e^{-i\varphi}|\uparrow\uparrow\rangle+\cos(\frac{\theta}{2}-\frac{\pi}{4})|\downarrow\downarrow\rangle,
\end{aligned}
\end{equation}
with the 2-qubit entangled degree $\mathcal{C}=|\cos\theta|$. 

The Berry phases for the eigenstates $|\psi_\pm\rangle$ can then be obtained~\cite{chen2007braiding}
\begin{equation}
\begin{aligned}\label{BE}
\gamma_\pm=i\int ^{2\pi}_{0}d\varphi \langle\psi_\pm|\frac{\partial}{\partial \varphi}|\psi_\pm\rangle=\pi (1\pm\sin\theta)=\pi (1\pm \sqrt{1-\mathcal{C}^2}).
\end{aligned}
\end{equation}
equation (~\ref{BE}) shows that the entangled degree of the system influences its Berry phase. Hence the Berry phase and entanglement are well connected in such a  simple Yang-Baxter system.


\subsection{YBE in topological basis, Wigner D-function $D^{j}(\theta,\varphi)$ and the role of $\ell_1$-norm}
 
 In this subsection, by means of topological fusion bases, we  reduce the 8D tensor form of YBE into the lower 2D dimension, where the solution of YBE usually appears as Wigner $D^{1/2}$-function. Based on such solution, we discuss the role of $\ell_1$-norm in YBE. 

%


Let us first introduce the Wigner $D^{j}$ function and its connections with the YBE. Any spin coherent operator, say $D(\theta,\phi)=e^{\xi J_+-\xi* J_-}$~\cite{Perelomov1977} is identical with the Euler rotation
\begin{equation}
D(\theta,\phi)=e^{i\phi J_z}e^{i2\theta J_y}e^{-i\phi J_z},
\end{equation}
where $J_x$, $J_y$, $J_z$ are $su(2)$ operators, obeying $[J_i,J_j]=i\epsilon_{ijk}J_k$, $\epsilon_{ijk}$ the Levy-Civita symbol. 
 
Then we can define the following  $D$-functions to satisfy YBE~\cite{niu2011role}
\begin{equation}
D(\theta_1,0)D(\theta_2,\phi)D(\theta_3,0)=D(\theta_3,\phi)D(\theta_2,0)D(\theta_1,\phi),
\end{equation}
with the constraint
\begin{equation}\label{p123}
\cos\phi=\frac{1}{2}\left[\frac{(\tan\theta_1+\tan\theta_3)-\tan\theta_2}{\tan\theta_1\tan\theta_2\tan\theta_3}-1\right].
\end{equation}
Here we note that the parameter relation can be directly derived from the $su(2)$ algebraic relation, i.e. independent of the concrete representations of $D$-function. 

When $\theta_1=\theta_2=\theta_3=\theta$, YBE reduces to the braid relation, which has been  already  proposed in reference~\cite{benvegnu2006uncertainty}, with the parameter relation
\begin{equation}
 \cos\phi=\frac{\cos2\theta}{1-\cos2\theta}.
\end{equation}


Now let us make connections between the $4\times4$ $\breve{R}$-matrix and the 2D Wigner $D^{1/2}$-function. 

For our $4\times4$ type-II $\breve{R}(\theta)$, the 8D YBE can be reduced  into a 2D space, spanned by two 4-qubit bases $\{|e_1\rangle,\, |e_2\rangle\}$~\cite{niu2011role},
\begin{align}
&|e_1\rangle=\frac{1}{\sqrt{2}}(|\psi_{12}\rangle|\psi_{34}\rangle+|\phi_{12}\rangle|\phi_{34}\rangle),\\
&|e_2\rangle=\frac{1}{\sqrt{2}}\left[(1+ie^{i\varphi})|\psi_{23}\rangle|\psi_{41}\rangle-(1-ie^{i\varphi})|\phi_{23}\rangle|\phi_{41}\rangle-|e_1\rangle\right],
\end{align}
where
\begin{equation}\label{EntangledSpin}
|\psi_{i,j}\rangle=\frac{1}{\sqrt{2}}\left(|\underset{i}{\uparrow}\underset{j}{\uparrow}\rangle+e^{-i\varphi}|\underset{i}{\downarrow}\underset{j}{\downarrow}\rangle\right), \, |\phi_{i,j}\rangle=\frac{1}{\sqrt{2}}\left(|\underset{i}{\uparrow}\underset{j}{\downarrow}\rangle-|\underset{i}{\downarrow}\underset{j}{\uparrow}\rangle\right),
\end{equation}
are Bell bases on the $i,j$ lattices. 
 Indeed, the two 4-qubit states can be regarded as the spin-$\frac{1}{2}$ realization of topological fusion bases (quantum dimension $d=\sqrt{2}$), whose graphical forms are expressed as follows 
 \begin{equation}\label{topological basis}
\begin{split}
   &|e_1\rangle =  \frac{1}{d}\,\, \begin{tikzpicture}[baseline=-\dimexpr\fontdimen22\textfont2\relax]
\draw[gray,line width=1.0pt] (-0.1,-0.1)--(-0.1,0.4) (-0.7,-0.1)--(-0.7,0.4) (0.1,-0.1)--(0.1,0.4) (0.7,-0.1)--(0.7,0.4);
\draw[gray,line width=1.0pt] (-0.7,-0.1) arc(180:360:0.3) (0.1,-0.1) arc(180:360:0.3);
\end{tikzpicture}\, ,\\
   &|e_2\rangle = \frac{1}{\sqrt{d^2-1}}\left(\,\, \begin{tikzpicture}[baseline=-\dimexpr\fontdimen22\textfont2\relax]
\draw[gray,line width=1.0pt] (-0.3,-0.2)--(-0.3,0.4) (-0.7,-0.3)--(-0.7,0.4) (0.3,-0.2)--(0.3,0.4) (0.7,-0.3)--(0.7,0.4);
\draw[gray,line width=1.0pt] (-0.3,-0.19) arc(240:300:0.6) (-0.7,-0.29) arc(240:300:1.4);
\end{tikzpicture} \,\,  - \frac{1}{d} \,\, \begin{tikzpicture}[baseline=-\dimexpr\fontdimen22\textfont2\relax]
\draw[gray,line width=1.0pt] (-0.1,-0.1)--(-0.1,0.4) (-0.7,-0.1)--(-0.7,0.4) (0.1,-0.1)--(0.1,0.4) (0.7,-0.1)--(0.7,0.4);
\draw[gray,line width=1.0pt] (-0.7,-0.1) arc(180:360:0.3) (0.1,-0.1) arc(180:360:0.3);
\end{tikzpicture}\,\, \right).
\end{split}
\end{equation}
The above two bases describe the processing of 4 Ising anyons fusing to the vacuum. They are nothing but the  1-qubit sparse encoding for Ising anyon in topological quantum computing.  The cap-cup form here is equivalent to the fusion tree  under Jones--Wenzl idempotent.
 
Applying the type-II solution $\breve{R}_{12}$, $\breve{R}_{23}$ onto the bases $\{|e_1\rangle,\, |e_2\rangle\}$, one can easily obtain the  2D YBE, as
\begin{equation}
\breve{R}_{12}(\theta)\longrightarrow A(\theta,\phi),\quad \breve{R}_{23}(\theta)\longrightarrow B(\theta,\phi),
\end{equation}
where
\begin{equation}
A(\theta)=V^\dag {D^{\frac{1}{2}}(\theta,\phi=0)} V, \,
B(\theta)=V^\dag {D^{\frac{1}{2}}(\theta,\phi=\frac{\pi}{2}) }V.
\end{equation}
Here $V=\frac{1}{\sqrt{2}}\left[\begin{matrix}1 & i \\ i & 1\end{matrix}\right]$, and $D^{\frac{1}{2}}(\theta,\phi)=\left[\begin{matrix}\cos\theta & -\sin\theta e^{-i\phi}\\ \sin\theta e^{i\phi}&\cos\theta\end{matrix}\right]$ is Wigner $D^{j}$-function for $j=1/2$.

More concretely, the 8D type-II YBE has 2D equivalence with $\phi=\frac{\pi}{2}$
\begin{equation}
A(\theta_1)B(\theta_2,\phi=\frac{\pi}{2})A(\theta_3)=B(\theta_3,\phi=\frac{\pi}{2})A(\theta_2)B(\theta_1,\phi=\frac{\pi}{2}).
\end{equation}
When $\phi=\frac{\pi}{2}$, according to equation (~\ref{p123}), we have the parameter constraint ($\phi=\frac{\pi}{2}$)
\begin{equation}
\tan\theta_2=\frac{\tan{\theta_1}+\tan{\theta_3}}{1+\tan{\theta_1}\tan{\theta_3}}.
\end{equation}
This Lorentzian additivity, coincides with the parameter relation for our type-II solution. 
When $\theta_1=\theta_2=\theta_3=\frac{\pi}{4}$, the 2D $\breve{R}$-matrix turns to the braid matrix associated to Ising anyon system~\cite{Nayak1996}
\begin{equation}
\mathcal{A}=A(\frac{\pi}{4})=\left[\begin{matrix}e^{-i\frac{\pi}{4}}& 0\\0 & e^{i\frac{\pi}{4}}\end{matrix}\right],\quad \mathcal{B}=B(\frac{\pi}{4})=\frac{1}{\sqrt{2}} \left[\begin{matrix}1& i\\ i &1\end{matrix}\right].
\end{equation}
And they obey braid relation
\begin{equation}
\mathcal{ABA}=\mathcal{BAB}.
\end{equation}

In our previous papers, we have shown that both the type-I and type-II $4\times4$ $\breve{R}$-matrices can be reduced to the 2D matrices. Then questions arise: What leads to the type-II(type-I) braid matrix $\mathcal{A}$ and $\mathcal{B}$ among the $D^{\frac{1}{2}}(\theta,\varphi)$? And why does $\theta=\frac{\pi}{4}$ lead to the maximum of entanglement?
It can be explained by  the maximum(minimum) of $\ell_1$-norm of $D^{\frac{1}{2}}(\theta,\varphi)$~\cite{niu2011role,GE2012,Ge2014Yangbaxter}.  

The applications of $\ell_1$-norm in quantum physics have already been proposed in recent years, such as in quantum process tomography~\cite{kosut2008quantum}, YBE~\cite{niu2011role} and quantifying coherence~\cite{baumgratz2014quantifying}, {\em et al.}   Usually in quantum mechanics, a wave function $|\Phi\rangle$ can be expanded as $|\Phi\rangle=\sum_i\alpha_i|\phi_i\rangle$, where $|\phi_i\rangle$ is orthonormal basis. The normalization of $|\Phi\rangle$ reads
\begin{equation}
\left\langle\Phi|\Phi\right\rangle=\sum_i|\alpha_i|^2=1.
\end{equation}  
We call $\sum_i|\alpha_i|^2=||\alpha||_{\ell_2}$ as $\ell_2$-norm, which indicates the square integrability or the probability distribution of the wave function.  Meanwhile, $\ell_1$-norm is defined as 
\begin{equation}\label{l1norm}
||\alpha||_{\ell_1}=\sum_i|\alpha_i|.
\end{equation}
 The minimization of $\ell_1$-norm plays important roles in information theory such as Compressed Sensing theory~\cite{donoho2006compressed,candes2006near}, {\em et al.} Hence, it is worthy of investigating whether the extremization of $\ell_1$-norm can be used to determine some important physical quantities in quantum information or in quantum mechanics.
%
%
%
%
%
%
%
%

%
%

For $D^{\frac{1}{2}}(\theta,\varphi)=\left[\begin{matrix}\cos\theta & -\sin\theta e^{-i\phi}\\ \sin\theta e^{i\phi}&\cos\theta\end{matrix}\right]$, the braid relation takes the form
\begin{equation}
D^{\frac{1}{2}}(\theta,0)D^{\frac{1}{2}}(\theta,\phi)D^{\frac{1}{2}}(\theta,0)=D^{\frac{1}{2}}(\theta,\phi)D^{\frac{1}{2}}(\theta,0)D^{\frac{1}{2}}(\theta,\phi).
\end{equation}
The constraint is
\begin{equation}
\cos\phi=\frac{\cos2\theta}{1-\cos2\theta}.
\end{equation}
Let $|\Psi\rangle=D^\frac{1}{2}|e_1\rangle=\cos\theta|e_1\rangle+\sin\theta e^{i\phi}|e_2\rangle$, then we can define the $\ell_1$-norm of state $|\Psi\rangle$
\begin{equation}
\||\Psi\rangle\|_{\ell_1}=|\cos\theta|+|\sin\theta e^{i\phi}|.
\end{equation} 
%
%
%
In connection with information theory~\cite{donoho2006compressed,candes2006near}, we define the $\ell_1$-norm as  $\sum^{N}_{n=1}|c_n|$ for a state $|\psi\rangle=\sum^{N}_{n=1}c_n|\psi_n\rangle$.
For $D^{\frac{1}{2}}(\theta,\phi)$, it is
\begin{equation}
\|D^{\frac{1}{2}}(\theta,\phi)\|_{\ell_1}=|\cos\theta|+|\sin\theta e^{i\phi}|=|\cos\theta|+|\sin\theta|.
\end{equation}
In our previous paper, the  extremum of $\ell_1$-norm has been well connected with the YBE as well as braid relation. Here we list the main results. For details, see references~\cite{niu2011role,GE2012,Ge2014Yangbaxter}. 
\begin{itemize}
\item Minimum  $\|D^{\frac{1}{2}}(\theta,\phi)\|_{\ell_1}$, $\theta=\frac{\pi}{2}$, $\phi=\frac{2\pi}{3}$, type-I braid matrix. YBE obeys Galilean additivity $\tan\theta_2=\tan\theta_1+\tan\theta_3$.
\item Maximum $\|D^{\frac{1}{2}}(\theta,\phi)\|_{\ell_1}$, $\theta=\frac{\pi}{4}$, $\phi=\frac{\pi}{2}$,  type-II braid matrix. YBE obeys Lorentzian additivity $\tan\theta_2=\frac{\tan\theta_1+\tan\theta_3}{1+\tan\theta_1\tan\theta_3}$. Such a maximum coincides with the maximum of von Neumann entropy. 
%
%
\end{itemize}
 
 In summary, the extremum of $\ell_1$-norm of Wigner $D$-function leads to two types of $2\times2$ braid matrices, which have real physical meanings among the solutions of YBE. The type-I $2\times2$ braid matrix and its $4\times4$ 6-vertex correspondence are connected to the traditional exactly solvable models, while the type-II $2\times2$ braid matrix and its $4\times4$ 8-vertex correspondence are related to the Ising anyon model and quantum entangled states. 

\subsection{Yang-Baxter equation in Majorana representation}

In previous sections, we mainly deal with the matrix form solution of YBE, which can be combined by the tensor of Pauli matrices.  In this subsection, we focus on the Majorana representation~\cite{ivanov2001non} of type-II Yang-Baxter solution~\cite{yu2015more}.

 Under Jordan-Wigner transformation, the Pauli matrices can be transformed into   the Majorana operators, as
\begin{equation}\label{MF}
\gamma_{j,A}=\left[\prod_{k=1}^{j-1}\sigma_k^z\right]\sigma_j^x, \quad \gamma_{j,B}=\left[\prod_{k=1}^{j-1}\sigma_k^z\right]\sigma_j^y.
\end{equation}
It is obvious that  one spin site corresponds to  two Majorana sites $\gamma_{A,B}$. And $\gamma_{i}$ obey Clifford algebra
\begin{equation}
\gamma_i=\gamma_i^{\dag}, \quad \{\gamma_{i,A}\, \gamma_{j,B}\}=2\delta_{ij}\delta_{AB}.
\end{equation}
Then the $\breve{R}$-matrix has the following equivalent deformation  
\begin{equation}
\breve{R}_{i,i+1}(\theta)=e^{-i\theta[\sigma^y\otimes\sigma^x]_{i,i+1}}=e^{{\theta\gamma_{i,A}\gamma_{i+1,A}}}.
\end{equation}
Here we note that the nearest neighbor Pauli representation is equivalent to the  next nearest neighbor Majorana representation.


We can also write out the nearest Majorana representation of $\breve{R}$, as follows
\begin{equation}
\begin{aligned}
&{\rm Odd-Even}:\quad \breve{R}_{2i-1,2i}(\theta)=e^{\theta \gamma_{i,A}\gamma_{i,B}},  \\
&{\rm Even-Odd}:\quad \breve{R}_{2i,2i+1}(\theta)=e^{\theta \gamma_{i,B}\gamma_{i+1,A}}. 
\end{aligned}
\end{equation}
As one can see from the above formulas, the nearest neighbor Majorana representation is subtly different from the next nearest neighbor Majorana representation: The odd-even $\breve{R}$ describes the Majorana $\gamma_i$'s interaction in one Pauli site, while the even-odd  $\breve{R}$ describes the Majorana $\gamma_i$'s interaction in two Pauli sites. Hence in the Pauli version, the odd-even and even-odd $\breve{R}$'s describe two different types of transverse Ising interactions. This is why the nearest neighbor Majorana representation of $\breve{R}$ leads to the Kitaev chain with topological phase. 

%
%
Now let us construct the 1D Kitaev chain~\cite{kitaev2001unpaired} based on the solution $\breve{R}_i(\theta)$ of YBE. We imagine that a unitary evolution is governed by $\breve{R}_i(\theta)$. Supposing that only $\theta$ in  $\breve{R}_i(\theta)$ is time-dependent, we can express a state $|\psi(t)\rangle$ as $|\psi(t)\rangle=\breve{R}_i(\theta(t))|\psi(0)\rangle$. Taking the Schr\"{o}dinger equation $i\hbar\tfrac{\partial}{\partial t}|\psi(t)\rangle=\hat{H}(t)|\psi(t)\rangle$ into account, one obtains:
\begin{equation}
i\hbar\tfrac{\partial}{\partial t}[\breve{R}_i|\psi(0)\rangle]=\hat{H}(t)\breve{R}_i|\psi(0)\rangle.
\end{equation}
Then the Hamiltonian $\hat{H}_i(t)$ related to the unitary operator $\breve{R}_i(\theta)$ is obtained:
\begin{equation}\label{SchrodingerEquation}
\hat{H}_i(t)=\textrm{i}\hbar\tfrac{\partial\breve{R}_i}{\partial t}\breve{R}_{i}^{-1}.
\end{equation}
Substituting $\breve{R}_i(\theta)=\exp(\theta\gamma_i\gamma_{i+1})$ into equation (~\ref{SchrodingerEquation}), we have:
\begin{equation}\label{2MFHamiltonian}
\hat{H}_i(t)=\textrm{i}\hbar\dot{\theta}\gamma_{i}\gamma_{i+1}.
\end{equation}
This Hamiltonian describes the interaction between $i$-th and $(i+1)$-th sites with the parameter $\dot{\theta}$. Indeed, when $\theta=n \times \tfrac{\pi}{4}$, the unitary evolution corresponds to the braiding progress of two nearest Majorana fermion sites in the system, here $n$ is an integer and represents  the times of the braiding operation.

If we only consider the nearest-neighbour interactions between MFs and extend equation (~\ref{2MFHamiltonian}) to an inhomogeneous chain with 2N sites, the derived ``superconducting''  chain model is expressed as:
\begin{equation}
\hat{H}=\textrm{i}\hbar\sum_{k=1}^{N}(\dot{\theta}_1\gamma_{k,A}\gamma_{k,B}+\dot{\theta}_2\gamma_{k,B}\gamma_{k+1,A}),
\end{equation}
with $\dot{\theta}_1$  and $\dot{\theta}_2$ describing odd-even and even-odd pairs, respectively.

Now we give a brief discussion about the above chain model in two cases:

\begin{enumerate}
\item $\dot{\theta}_1>0$, $\dot{\theta}_2=0.$

In this case, the Hamiltonian is:
\begin{equation}\label{YBEtrivil}
\hat{H}_1=\textrm{i}\hbar\sum_{k}^{N}\dot{\theta}_1\gamma_{k,A}\gamma_{k,B}.
\end{equation}
As defined in equation (~\ref{MF}), the Majorana operators $\gamma_{k,A}$ and $\gamma_{k,B}$ come from the same ordinary fermion site k, $\textrm{i}\gamma_{k,A}\gamma_{k,B}=2a_{k}^{\dag}a_{k}-1$ ($a_{k}^{\dag}$ and $a_{k}$ are spinless ordinary fermion operators). $\hat{H}_1$ simply means the total occupancy of ordinary fermions in the chain and has U(1) symmetry, $a_j\rightarrow e^{i\phi}a_j $.  Specifically, when $\theta_1(t)=\tfrac{\pi}{4}$, the unitary evolution $e^{\theta_{1}\gamma_{2k-1}\gamma_{2k}}$ corresponds to the braiding operation of two Majorana sites from the same k-th ordinary fermion site.  The ground state represents the ordinary fermion occupation number 0. In comparison to 1D Kitaev model, this Hamiltonian corresponds to the trivial case of Kitaev's. The unitary evolution of the system $e^{-i{\int\hat{H}_1dt}}$ stands for the exchange process of odd-even Majorana sites.

\item $\dot{\theta}_1=0$, $\dot{\theta}_2>0.$

In this case, the Hamiltonian is:
\begin{equation}\label{YBEtopo}
\hat{H}_2=\textrm{i}\hbar\sum_{k}^{N}\dot{\theta}_2\gamma_{k,B}\gamma_{k+1,A}.
\end{equation}
This Hamiltonian corresponds to the topological phase of 1D Kitaev model and has $\mathbb{Z}_2$ symmetry, $a_j\rightarrow -a_j$. Here the operators $\gamma_{1,A}$ and $\gamma_{N,B}$ are absent in $\hat{H}_2$.  The Hamiltonian has two degenerate ground state, $|0\rangle$ and $|1\rangle=d^{\dag}|0\rangle$, $d^{\dag}=e^{-i\varphi/2}(\gamma_{1,A}-i\gamma_{N,B})/2$. This mode is the so-called Majorana mode in 1D Kitaev chain model. When $\theta_2(t)=\tfrac{\pi}{4}$, the unitary evolution $e^{\theta_{2}\gamma_{k,B}\gamma_{k+1,A}}$ corresponds to the braiding operation of two Majorana sites $\gamma_{k,B}$ and $\gamma_{k+1,A}$ from $k$-th and $(k+1)$-th ordinary fermion sites, respectively.

\medskip
\end{enumerate}

Thus we conclude that our Hamiltonian derived from $\breve{R}_{i}(\theta(t))$ corresponding to the braiding of nearest Majorana fermion sites is exactly the same as the 1D wire proposed by Kitaev, and $\dot{\theta}_1=\dot{\theta}_2$ corresponds to the phase transition point in the ``superconducting'' chain. By choosing different time-dependent parameters $\theta_1$ and $\theta_2$, we find that the Hamiltonian $\hat{H}$ corresponds to different phases.

\subsection{ 3-body factorized S-matrix constrained by YBE}

Having obtained the properties of the 2-body scattering matrix $\breve{R}$, let us now discuss the factorized 3-body scattering matrix composed from the 2-body S-matrices via YBE. In S-matrix theory, the factorization means that if a 3-body scattering matrix $S$ can be decomposed to three 2-body S-matrices, then many-body $S$-matrix can be down.
The YBE provides a constraint for the factorized  3-body S-matrix
\begin{equation}
\breve{S}_{123}=\breve{R}_{12}(\theta_1) \breve{R}_{23}(\theta_2) \breve{R}_{12}(\theta_3)=\breve{R}_{23}(\theta_3) \breve{R}_{12}(\theta_2) \breve{R}_{23}(\theta_1).
\end{equation}
That is, there are two equivalent ways in decomposing 3-body S-matrix into three 2-body S-matrices. 
 
For the type-II solution of YBE, we have the following factorized 3-body S-matrix~\cite{yu2014factorized}
\begin{equation}\label{3MFExp}
  \breve{S}_{123}(\eta,\beta)=e^{\eta \left(\vec{n}\cdot\vec{\Lambda}\right)},
\end{equation}
where 
\begin{eqnarray*}
 \cos\eta &=& \cos\theta_2 \cos\left(\theta_1+\theta_3\right),\\
 \sin\eta &=& \sin\theta_2\sqrt{1+\cos^2(\theta_1-\theta_3)},\\
 \vec{n} &=& \left(
  \begin{array}{ccc}
  \tfrac{1}{\sqrt{2}}\cos\beta,& \tfrac{1}{\sqrt{2}}\cos\beta,& \sin\beta
  \end{array}\right),\\
  \vec{\Lambda} &=& (\gamma_{1,A}\gamma_{2,A},\, \gamma_{2,A}\gamma_{3,A},\,\gamma_{1,A}\gamma_{3,A}),\textrm{(still SU(2))}\\
  		&=&(-i\sigma^y\sigma^x I,\, -iI\sigma^y\sigma^x,\, -i\sigma^y\sigma^z\sigma^x),\\
  \cos\beta &=& \tfrac{\sqrt{2}\cos\left(\theta_1-\theta_3\right)}{\sqrt{1+\cos^2\left(\theta_1-\theta_3\right)}},\\               
  \sin\beta &=& \tfrac{-\sin\left(\theta_1-\theta_3\right)}{\sqrt{1+\cos^2\left(\theta_1-\theta_3\right)}}.
 \end{eqnarray*}
 It is easy to check that the sectors in $\vec{\Lambda}$ still obey the $su(2)$ algebraic relation. Hence the $ \breve{S}_{123}(\eta,\beta)$ can be regarded as the rotation in 3D space spanned by $su(2)$ generators. 
 
Following the Yang-Baxter solution $\breve{R}$ in describing 2-qubit entanglement, we find that $\breve{S}_{123}(\eta,\beta)$ also generates 3-qubit entanglement.  Significantly, $\breve{S}_{123}(\eta,\beta)$ is able to generate the two types of maximal 3-qubit entangled states, say, GHZ state and $W$ state.
\begin{itemize}
\item When $\eta=\frac{\pi}{3}, \, \beta=\textrm{arccot} \sqrt{2}$, 
\begin{equation}
\begin{aligned}
\breve{S}_{123}|000\rangle&=\frac{1}{2}(|000\rangle+|011\rangle+|101\rangle+|110\rangle)\\
&=\frac{1}{\sqrt{2}}(|+\rangle^{\otimes3}+|-\rangle^{\otimes3}), \,\textrm{(GHZ state)}.
\end{aligned}
\end{equation}
\item When $\eta=\frac{\pi}{2}, \, \beta=\textrm{arccot} \sqrt{2}$, 
\begin{equation}
\breve{S}_{123}|000\rangle=\frac{1}{\sqrt{3}}(|011\rangle+|101\rangle+|110\rangle),\,\textrm{(W state)}.
\end{equation}
\end{itemize} 
It is well known that in 3-qubit pure state, under SLOCC, GHZ state and $W$ state are the only two types of non-equivalent genuine entangled states. Hence our 3-body $S$-matrix describes all 3-qubit maximal entangled pure states in this sense.

We suppose that the parameter $\eta$ is time-dependent and $\beta$ is time-independent in $\breve{S}_{123}(\eta,\beta)$, then the desired 3-body Hamiltonian can be obtained from equation (~\ref{SchrodingerEquation}):
\begin{eqnarray}\label{3bodyHamilton}
\hat{H}_{123}(t)&=&\textrm{i}\hbar\tfrac{\partial\breve{S}_{123}}{\partial t}\breve{S}_{123}^{-1}\nonumber\\
              &=&\textrm{i}\hbar\dot{\eta}\left[\tfrac{1}{\sqrt{2}}\cos\beta(\gamma_1\gamma_2+\gamma_2\gamma_3) +\sin\beta \gamma_1\gamma_3\right].
\end{eqnarray}
The constructed Hamiltonian, which has been mentioned in references~\cite{alicea2011non,lee2013algebra}, describes the 2-body interactions among the three Majorana operators. It describes the effective interaction in a $T$-junction formed by three quantum wires. In reference~\cite{lee2013algebra}, it has been shown that the above Hamiltonian, which commutes with emergent Majorana operator $\Gamma=-\textrm{i}\gamma_1\gamma_2\gamma_3$, holds Majorana doubling.  From the viewpoint of YBE,  the intrinsic commutation relation is between  $\Gamma$ and the solution of YBE $\breve{R}_i(\theta)=e^{\theta\gamma_i\gamma_{i+1}}$. It is shown that:
\begin{equation}
[\Gamma, \breve{R}_{i}(\theta)]=0 ,  \quad (i=1,2).
\end{equation}
Indeed, the above commutation relation indicates that emergent Majorana operator $\Gamma$ is a new symmetry of the solution $\breve{R}_i(\theta)$ of YBE. It is due to the decomposition of 3-body interaction into three 2-body interactions via the approach of YBE that the derived Hamiltonian holds Majorana doubling.

The  extended emergent Majorana mode $\Gamma'$ supporting odd number $N$ of Majorana operators~\cite{lee2013algebra} is,
\begin{equation}
\Gamma'\equiv\textrm{i}^{N(N-1)/2}\prod_{j=1}^{N}\gamma_j .
\end{equation}
It is easy to check that:
\begin{equation}
\left[\Gamma', B_i\right]=0, \quad (i=1,2,...N-1),
\end{equation}
where $B_i=e^{\tfrac{\pi}{4}\gamma_i\gamma_{i+1}}$ is the generator of the braid group $B_N$. The commutation relation indicates that $\Gamma'$ plays the role of an invariant in the braid group $B_N$, we call $\Gamma'$ the collective discrete symmetry.
%

\section{$9\times9$ Solution of Yang-Baxter Equation}\label{sec3}
In this section, we will generalize the $4\times4$ case into the $9\times9$ case. Let us first present the $9\times9$ braid matrix and its Yang-Baxterization. 

We adopt the following $9\times9$ braid matrix~\cite{Rowell2009localization},
\begin{equation}
B=\frac{\textrm{i}}{\sqrt{3}}\left[\begin{array}{ccccccccc}
\omega & 0 & 0 & 0 & 0 & 1 & 0 & \omega & 0\\
0 & \omega & 0 & \omega^2 & 0 & 0 & 0 & 0 & \omega^2\\
0 & 0 & \omega & 0 & \omega & 0 & 1 & 0 & 0\\
0 & \omega^2 & 0 & \omega & 0 & 0 & 0 & 0 & \omega^2\\ 
0& 0 & 1 & 0 & \omega & 0 & \omega & 0 & 0\\
\omega & 0 & 0 & 0 & 0 & \omega & 0 & 1 & 0\\
0 & 0 & \omega & 0 & 1 & 0 & \omega & 0 & 0\\
1 & 0 & 0 & 0 & 0 & \omega & 0 & \omega & 0\\
0 & \omega^2 & 0 & \omega^2 & 0 & 0 & 0 & 0 & \omega 
\end{array}\right],
\end{equation}
with $\omega=e^{i2\pi/3}$, and it obeys braid relation
\begin{equation}
(B\otimes \mathbb{I}_3)(\mathbb{I}_3\otimes B)(B\otimes \mathbb{I}_3)=(\mathbb{I}_3\otimes B)(B\otimes \mathbb{I}_3)(\mathbb{I}_3\otimes B).
\end{equation}
 
Similar to the Bell braid matrix, if applying $B$ on 2-qutrit tensor bases, one also obtains the 2-qutrit entangled states:
\begin{equation}
B\left[\begin{array}{l}|00\rangle\\|01\rangle\\|02\rangle\\|10\rangle\\|11\rangle\\|12\rangle\\|20\rangle\\|21\rangle\\|22\rangle\end{array}\right]
=\frac{i}{\sqrt{3}}\left[\begin{array}{l}
\omega|00\rangle+|12\rangle+\omega|21\rangle\\ 
\omega|01\rangle+\bar{\omega}|10\rangle+\bar{\omega}|22\rangle\\
\omega|02\rangle+\omega|11\rangle+|20\rangle\\
\omega|10\rangle+\bar{\omega}|01\rangle+\bar{\omega}|22\rangle\\
\omega|11\rangle+|02\rangle+\omega|20\rangle\\
\omega|12\rangle+\omega|00\rangle+|21\rangle\\
\omega|20\rangle+\omega|02\rangle+|11\rangle\\
\omega|21\rangle+|00\rangle+\omega|12\rangle\\
\omega|22\rangle+\bar{\omega}|01\rangle+\bar{\omega}|10\rangle
\end{array}\right].
\end{equation}
 This braid representation is associated to Temperley-Lieb algebra with quantum dimension $d=\sqrt{3}$, hence we can make the simplest rational Yang-Baxterization, which gives~\cite{yu2016z3} 
 \begin{equation}
\begin{aligned}
&{\breve{R}(\theta)=e^{i\theta M}=\cos\theta*(\mathbb{I}_9+\tan\theta M),}\\
&M=-\frac{1}{\sqrt{3}}(\mathbb{I}_9+2e^{i\frac{2\pi}{3}}B),\, M^2=-\mathbb{I}_9.
\end{aligned}
\end{equation}
It seems like trigonometric Yang-Baxterization solution, but if we set $u=\tan\theta$, the solution turns into the rational form. To satisfy YBE $\breve{R}_{12}(\theta_1) \breve{R}_{23}(\theta_2) \breve{R}_{12}(\theta_3)=\breve{R}_{23}(\theta_3) \breve{R}_{12}(\theta_2) \breve{R}_{23}(\theta_1)$, the constraint reads
\begin{equation}\label{3Lorentz}
{\tan\theta_2=\frac{\tan\theta_1+\tan\theta_3}{1+\frac{1}{3}\tan\theta_1\tan\theta_3}.}
\end{equation} 
We see from equation (~\ref{3Lorentz}) that when  $u=\frac{1}{\sqrt{3}}\tan\theta$, the relation becomes the Lorentzian type.   This is similar to the $4\times4$ case. Applying $9\times9$ $\breve{R}$-matrix on the natural basis also leads to the 2-qutrit entangled state, with the entangled degree determined by $\theta$. 

In the following subsections, we will introduce some properties of the $9\times9$ $\breve{R}$-matrix.

\subsection{$\mathbb{Z}_3$ parafermion version of $\breve{R}$-matrix and Fendley chain}
In this subsection, we mainly review the $9\times9$ matrix representation and parafermionic representation of Yang-Baxter solution.

Similar to the relationship between the Pauli matrices and Majorana fermions via Jordan-Wigner transformation, 
the Fradkin-Kadanoff transformation~\cite{fradkin1980disorder} connects the principal representation of $SU(3)$ with the $\mathbb{Z}_3$ parafermions, as
\begin{equation}\label{z3jw}
\begin{aligned}
&C_{\textrm{2k-1}}=(Z^{\dagger})^{\otimes (k-1)}\otimes X^{\dagger}_{\textrm{k}}\otimes I^{\otimes (N-k)},\\
&C_{\textrm{2k}}=(Z^{\dagger})^{\otimes (k-1)}\otimes(XZ)^{\dagger}_{\textrm{k}}\otimes I^{\otimes (N-k)},
\end{aligned}
\end{equation}
where
\begin{equation}
Z=\left[\begin{array}{ccc} 1 & 0 & 0 \\ 0 & \omega & 0 \\ 0 & 0 & \omega^2 \end{array}\right], \, X=\left[\begin{array}{ccc} 0 & 1 & 0 \\ 0 & 0 & 1 \\ 1 & 0 & 0 \end{array}\right], 
\end{equation}
are part of generators of $SU(3)$ in Kac principal representation~\cite{kac1994infinite}.  $C_i$'s are 
$\mathbb{Z}_3$ parafermions and obey Heisenberg-Weyl algebra
$$(C_i)^3=1,\, C_iC_j=\omega^{\textrm{sgn}|j-i|}C_jC_i, \, \omega=e^{\text{i}\frac{2\pi}{3}}.$$ 
 
Here we directly discuss the nearest neighbor parafermion representation of $\breve{R}$.  The solution $\breve{R}(\theta)$ takes the following form,
 \begin{equation}\label{CrepR}
\breve{R}_{\textrm{i}}(\theta)\rightarrow e^{-\textrm{i}\theta}+\textrm{i}\frac{2}{3}\sin\theta\left[1+\omega^2C_{\textrm{i}}^{\dagger}C_{\textrm{i}+1}+\omega^2C_{\textrm{i}}C_{\textrm{i}+1}^{\dagger}\right],
\end{equation}
with the same constraint:
\begin{equation}\label{ParaYBEAngRel}
\tan\theta_2=\frac{\tan\theta_1+\tan\theta_3}{1+\frac{1}{3}\tan\theta_1\tan\theta_3}.
\end{equation}

Now let us construct the $\mathbb{Z}_3$ parafermionic chain based on equation (~\ref{CrepR}). Substituting equation (~\ref{CrepR}) into equation (~\ref{SchrodingerEquation}), we get
\begin{equation}\label{2CHamiltonian}
\hat{H}_{\textrm{i}}(t)=-\dot{\theta}\frac{2\hbar}{3}\left(\omega^2C_{\textrm{i}}^{\dagger}C_{\textrm{i}+1}+\omega^2C_{\textrm{i}}C_{\textrm{i}+1}^{\dagger}-\frac{1}{2}\right).
\end{equation}
Similarly, we consider the nearest-neighbour interactions of $C_{\textrm{i}}$'s and extend equation (~\ref{2CHamiltonian}) to an 2N-site chain  and ignore the constant term, the derived chain model can be expressed as:
\begin{equation}\label{YBEcolorKitaev}
\hat{H}=-\frac{2\hbar}{3}\omega^2{\Big [ }\sum_{i=1}^{N}\dot{\theta}_1\left(C_{\textrm{2i}-1}^{\dagger}C_{\textrm{2i}}+C_{\textrm{2i}-1}C_{\textrm{2i}}^{\dagger}\right)+\sum_{i=1}^{N-1}\dot{\theta}_2\left(C_{\textrm{2i}}^{\dagger}C_{\textrm{2i}+1}+C_{\textrm{2i}}C_{\textrm{2i}+1}^{\dagger}\right)\Big].
\end{equation}
 Here we emphasize that the chain possesses open boundary condition. This model is the 1D $\mathbb{Z}_3$ parafermionic model~\cite{fendley2012parafermionic} with chiral symmetry, which originates from the three-state Potts model~\cite{baxter1988new,stephen1972pseudo,fateev1982self}. Instead of the $\mathbb{Z}_2$ parity symmetry of Kitaev model, the model in equation (~\ref{YBEcolorKitaev}) possesses $\mathbb{Z}_3$ symmetry.  The symmetry operator is 
\begin{equation}
P=\prod_{\textrm{i}=1}^{\textrm{N}}\left(C_{2\textrm{i}-1}^{\dagger}C_{2\textrm{i}}\right), \quad P^3=1.
\end{equation}
Hence $P$ is a $\mathbb{Z}_3$ symmetry of the model and the eigenvalues of $P$ is $1$, $\omega$ and $\omega^2$. The exact solubility of this model has been discussed in previous papers~\cite{fendley2012parafermionic,fendley2014free,alicea2016topological}.  Here we only  analyze the obtained model in two simple cases.
\begin{enumerate}
\item $\dot{\theta}_1>0$, $\dot{\theta}_2=0.$

In this case the Hamiltonian becomes:
\begin{eqnarray}
\hat{H}_1&=&-\frac{2\hbar}{3}\dot{\theta}_1\omega^2\sum_{i=1}^{N}\left(C_{\textrm{2i}-1}^{\dagger}C_{\textrm{2i}}+C_{\textrm{2i}-1}C_{\textrm{2i}}^{\dagger}\right)\nonumber\\
&=&\frac{2\hbar}{3}\dot{\theta}_1\sum_{i=1}^{N}\left[d_{\textrm{i}}^{\dagger}d_{\textrm{i}}-2\right].\label{YBESU3trivil}
\end{eqnarray}
Here we note that $C_{\textrm{2i}-1}$ and $C_{\textrm{2i}}$ correspond to i-th SU(3) spin, $d_{\textrm{i}}^{\dagger}=C_{\textrm{2i-1}}^{\dag}-\omega C_{\textrm{2i}}^{\dag}$, and the vacuum state $|0\rangle$ is defined as $d_{\textrm{i}}|0\rangle=0$. The Hamiltonian is diagonalized and  the ground state is unique. This is a trivial case. 
 
\item $\dot{\theta}_1=0$, $\dot{\theta}_2>0.$

In this case the Hamiltonian is:
\begin{eqnarray}
\hat{H}_2&=&-\frac{2\hbar}{3}\dot{\theta}_2\omega^2\sum_{k=1}^{N-1}\left(C_{\textrm{2i}}^{\dagger}C_{\textrm{2i}+1}+C_{\textrm{2i}}C_{\textrm{2i}+1}^{\dagger}\right)\\
&=&\frac{2\hbar}{3}\dot{\theta}_2\sum_{i=1}^{N-1}\left[\tilde{d}_{\textrm{i}}^{\dagger}\tilde{d}_{\textrm{i}}-2\right].\label{YBESU3topo}
\end{eqnarray}
Here the quasiparticle at lattice can be defined as $\tilde{d}_{\textrm{i}}=C_{\textrm{2i}}-\omega^2C_{\textrm{2i}+1}$. The ground states satisfy the condition $\tilde{d}_{\textrm{i}}|\psi\rangle=0$ for $i=1,..., N-1$.  Under the open boundary condition, it shows that the absent operators $C_{\textrm{1}}$, $C_{\textrm{1}}^{\dagger}$,  $C_{\textrm{2N}}$ and $C_{\textrm{2N}}^{\dagger}$ in $\hat{H}_2$ remain unpaired and are the symmetry operators of the Hamiltonian $\hat{H}_2$. Together with the $\omega$-parity operator $P$,  $C_{\textrm{1}}$, $C_{\textrm{1}}^{\dagger}$,  $C_{\textrm{2N}}$ and $C_{\textrm{2N}}^{\dagger}$ lead to the triple degeneracy of ground states which can be categorized according to $P$.  The Hamiltonian has three degenerate ground states: $|\psi_0\rangle$, $|\psi_1\rangle=d^{\dag}|\psi_0\rangle$ and $|\psi_2\rangle=[d^{\dag}]^2|\psi_0\rangle$, where  $d^{\dag}=C_{\textrm{1}}^{\dag}-\omega C_{\textrm{2N}}^{\dag}$. The three ground states $|\psi_0\rangle$, $|\psi_1\rangle$ and $|\psi_2\rangle$ possess the parity $1$, $\omega$ and $\omega^2$, respectively.
\medskip
\end{enumerate}

Indeed, the critical point of the phase transition $|\dot{\theta}_2|=|\dot{\theta}_1|$ in the Hamiltonian (~\ref{YBEcolorKitaev}) coincides with the $\mathbb{Z}_3$ conformal field theory(CFT)~\cite{dotsenko1984critical,zamolodchikov1985nonlocal,mong2014parafermionic}. Obviously, the above properties in our derived $\mathbb{Z}_3$ parafermionic chain are very similar to 1D Kitaev model. However, there are still some differences between  $\mathbb{Z}_2$ and $\mathbb{Z}_3$ models. The critical point of $\mathbb{Z}_2$ Kitaev model can be described by Ising CFT. When Kitaev model is in topological phase, it appears Majorana zero mode with quantum dimension $\sqrt{2}$. While the critical point of $\mathbb{Z}_3$ parafermion model is also described by $\mathbb{Z}_3$ parafermion CFT, but the non-abelian primary fields are not $\mathbb{Z}_3$ parafermion field. There are totally six different quasiparticles in $\mathbb{Z}_3$ parafermion model, three of which possess abelian fields, the vacuum 1, parafermion field $\psi$ and $\psi^{\dag}$. Besides, there exist three types of non-abelian fields, $\sigma$, $\sigma^\dag$, $\epsilon$, where $\sigma$ is the spin field and $\epsilon=\sigma\psi$  is the Fibonacci anyon with quantum qimension $\frac{1+\sqrt{5}}{2}$. For example, the $\mathbb{Z}_3$ Read-Rezayi quantum Hall phase supports Fibonacci anyons, which are applicable to universal quantum computation~\cite{mong2014parafermionic,read1999beyond,nayak2008non}.
\subsection{$\ell_1$-norm in $9\times9$ Yang-Baxter system}
In the previous section, the role of $\ell_1$-norm in $4\times4$ Yang-Baxter system has been presented. Here we extend the results to the $9\times9$ Yang-Baxter system~\cite{Yu2017d}. For nearest neighbor parafermion representation, the braid matrix in 2-qutrit space $(\mathbb{C}_3)^{\otimes 2}$  is (Basis:$\{|1\rangle,|2\rangle,|3\rangle\}^{\otimes2}$)
\begin{eqnarray}
&&B_{12}=B_{1}=\left[\begin{array}{ccc} e^{-i\frac{\pi}{3}} & 0 & 0 \\ 0 &e^{i\frac{\pi}{3}} & 0 \\ 0 & 0 &e^{-i\frac{\pi}{3}} \end{array}\right]\otimes\left[\begin{array}{ccc} 1 & 0 & 0 \\ 0 &1& 0 \\ 0 & 0 &1 \end{array}\right],\label{B12}\\
&&B_{23}=B_{2}=\frac{\textrm{i}\omega}{\sqrt{3}}\left[\begin{matrix}
\omega & 0 & 0 & 0 & 0 & \omega & 0 & 1 & 0\\
0 & \omega & 0 & \omega & 0 & 0 & 0 & 0 & 1\\
0 & 0 & \omega & 0 & \omega & 0 & 1 & 0 & 0\\
0 & 1 & 0 & \omega & 0 & 0 & 0 & 0 & \omega\\ 
0& 0 & 1 & 0 & \omega & 0 & \omega & 0 & 0\\
1 & 0 & 0 & 0 & 0 & \omega & 0 & \omega & 0\\
0 & 0 & \omega & 0 & 1 & 0 & \omega & 0 & 0\\
\omega & 0 & 0 & 0 & 0 & 1 & 0 & \omega & 0\\
0 & \omega & 0 & 1& 0 & 0 & 0 & 0 & \omega
\end{matrix}\right].\label{B23}
\end{eqnarray} 
Correspondingly, the parametrized $\breve{R}$-matrix takes the form $\breve{R}_{i}(\theta)=\frac{2}{\sqrt{3}}[\cos(\theta+\pi/6)+\sin\theta B_i].$

In this case,  the YBE lies in the $9\times9$ space spanned by 2-qutrit natural bases. Further, the $9\times9$ matrix can be reduced to $3\times3$ matrix,  under the conservation of 2-qutrit  parity operator $P$,
\begin{equation}\label{parity}
P^3=1,\, P|ij\rangle=\omega^{(i+j)}|ij\rangle,\quad \omega=\exp(i2\pi/3).
\end{equation}
We then have the commutation relations $[\breve{R}_{12},P]=[\breve{R}_{23},P]=0$, i.e. $\breve{R}$ preserves the parity of the system. Hence we can choose one subspace $\{|11\rangle, |23\rangle, |32\rangle\}$ where $P=\exp(i4\pi/3)$. Then the reduced $3\times3$ $\breve{R}$-matrix reads as follows
\begin{eqnarray}
&&\breve{R}_{12}\rightarrow\mathcal{A}_{12}(\theta)=\left[\begin{array}{ccc} e^{-i\theta} & 0 & 0 \\ 0 & e^{i\theta} & 0 \\ 0 & 0 & e^{-i\theta} \end{array}\right],  \\
&&\breve{R}_{23}\rightarrow\mathcal{A}_{23}(\theta)=\left[\begin{smallmatrix}\cos\theta-\frac{i}{3}\sin\theta &\frac{2i}{3}\omega^2\sin\theta& \frac{2i}{3}\omega\sin\theta \\ \frac{2i}{3}\omega\sin\theta & \cos\theta-\frac{i}{3}\sin\theta & \frac{2i}{3}\omega^2\sin\theta  \\ \frac{2i}{3}\omega^2\sin\theta & \frac{2i}{3}\omega\sin\theta & \cos\theta-\frac{i}{3}\sin\theta \end{smallmatrix}\right].             
\end{eqnarray}
The point here is that the $9\time9$ matrices can be decomposed into three 3D blocks, i.e. $3\otimes3=3\oplus3\oplus3$.
Easily we can check that $\mathcal{A}_{12}(\theta)$ and $\mathcal{A}_{23}(\theta)$ satisfy YBE 
\begin{equation}
\mathcal{A}_{12}(\theta_1)\mathcal{A}_{23}(\theta_2)\mathcal{A}_{12}(\theta_3)=\mathcal{A}_{23}(\theta_3)\mathcal{A}_{12}(\theta_2)\mathcal{A}_{23}(\theta_1),
\end{equation}
with the same constraint as the $\mathbb{Z}_3$ parafermionic representation,
\begin{equation}
\tan\theta_2=\frac{\tan\theta_1+\tan\theta_3}{1+\frac{1}{3}\tan\theta_1\tan\theta_3}.
\end{equation}

Having noticed the parity symmetry in YBE,  now let us discuss the entanglement properties in terms of the von Neumann entropy and $\ell_1$-norm in Yang-Baxter system. If applying $\mathcal{A}_{23}(\theta)$ on basis $|11\rangle$, one obtains the state in the subspace $\{|11\rangle, |23\rangle, |32\rangle\}$:
\begin{equation}
|\Psi(\theta)\rangle=(\cos\theta-\frac{i}{3}\sin\theta)|11\rangle+\frac{2i}{3}\omega^2\sin\theta|32\rangle+\frac{2i}{3}\omega\sin\theta|23\rangle.
\end{equation}
$\ell_1$-norm of the Yang-Baxter solution can be defined via $\|\Psi\|_{\ell_1}$, as
\begin{equation}\|\Psi\|_{\ell_1}=|\cos\theta-\frac{i}{3}\sin\theta|+|\frac{2i}{3}\omega^2\sin\theta|+|\frac{2i}{3}\omega\sin\theta|.
\end{equation}

\begin{figure}[h]
\includegraphics[width=22pc]{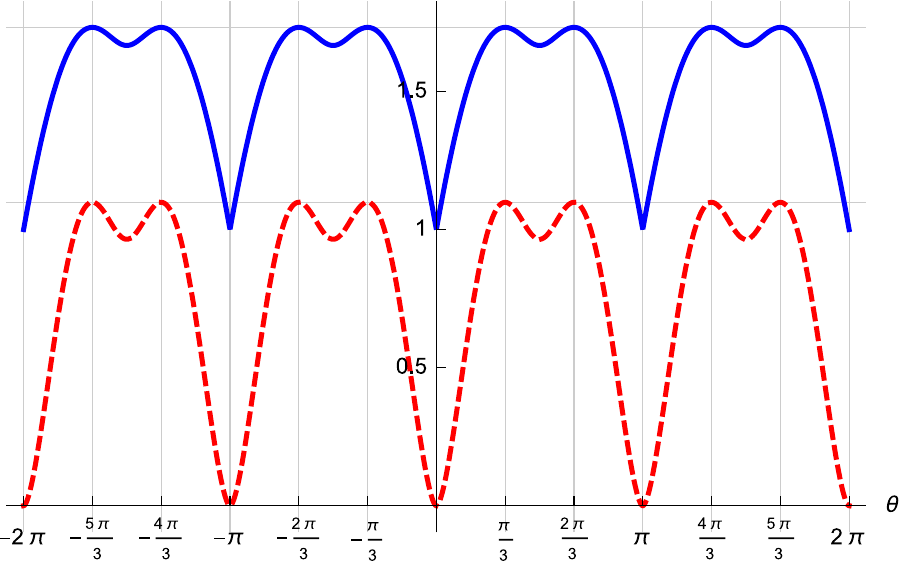}\hspace{2pc}%
\begin{minipage}[b]{14pc}\caption{\label{L1vN}Von Neumann entropy and $\ell_1$-norm of $|\Psi(\theta)\rangle$ as a function of $\theta$ that describes the qutrit entanglement. The von Neumann entropy is labeled by the red dashed line, and the $\ell_1$-norm is labeled by blue solid line. Both of them achieve the maximal extremum at $\theta=\frac{\pi}{3}$, at which value the $\breve{R}(\theta)$ reduces braid matrix.}
\end{minipage}
\end{figure}

On the other hand, by calculating the von Neumann entropy of the state $|\Psi\rangle$, $S(\rho)=-\textrm{tr}[\rho \ln\rho]$,
\begin{equation}
\begin{split}
S=&-|\cos\theta-\frac{i}{3}\sin\theta|^2\ln|\cos\theta-\frac{i}{3}\sin\theta|^2\\
&-|\frac{2i}{3}\sin\theta|^2\ln|\frac{2i}{3}\sin\theta|^2-|\frac{2i}{3}\sin\theta|^2\ln|\frac{2i}{3}\sin\theta|^2,
\end{split}
\end{equation}
we find that the extremum of $\ell_1$-norm coincide exactly with the extremum of von Neumann entropy, as shown in figure~\ref{L1vN}. The most interesting result is when the $\ell_1$-norm and von Neumann entropy take the maximum values, the Yang-Baxter solution $\breve{R}$ turns out to be the braid matrix, which has the real physical meaning.

In addition, we would like to introduce the discrete global symmetry of $9\times9$ YBE.

To satisfy YBE, 
\begin{equation}
\breve{R}_{12}(\theta_1) \breve{R}_{23}(\theta_2) \breve{R}_{12}(\theta_3)
=\breve{R}_{23}(\theta_3) \breve{R}_{12}(\theta_2) \breve{R}_{23}(\theta_1),
\end{equation}
the $\mathbb{Z}_3$ parafermionic solution takes the following form
\begin{equation}
\breve{R}_{\textrm{i,i+1}}(\theta)= e^{-\textrm{i}\theta}+\textrm{i}\frac{2}{3}\sin\theta\left[1+\omega^2C_{\textrm{i}}^{\dagger}C_{\textrm{i}+1}+\omega^2C_{\textrm{i}}C_{\textrm{i}+1}^{\dagger}\right].
\end{equation}

Then the YBE possesses the following collective discrete symmetry,
\begin{eqnarray*}
 && P=C_1C_2^{\dag}C_3C_4^{\dag} ,\textrm{(Parity)}\\
 && \Gamma=\omega C_1^{\dag}C_2C_3^{\dag}, \,\textrm{(Emergent parafermion mode)},\\
 && P^3=1,\, \Gamma^3=1,\, [P,\, \Gamma]\neq0.\\
  && [P,\, \breve{R}_{\textrm{i,i+1}}]=0,\, [\Gamma,\, \breve{R}_{\textrm{i,i+1}}]=0.
\end{eqnarray*}
 Extending to the  $(2n-1) $-parafermion system obtained from YBE, the system also hosts collective discrete symmetry, 
  \begin{eqnarray*}
 &&P=\prod_{i=1}^{n+1}C_{2i-1}C_{2i}^{\dag},\textrm{(Parity)}\\
 &&\Gamma=C_{2n+1}^{\dag}\prod_{i=1}^{n}C_{2i-1}^{\dag}C_{2i} ,\,\textrm{(Emergent parafermion mode)},\\
  && P^3=1,\, \Gamma^3=1,\, [P,\, \Gamma]\neq0.\end{eqnarray*}
This is analogous to the algebra of Majorana doubling proposed by Lee-Wilczek, we call it the algebra of $\mathbb{Z}_3$ parafermion triple degeneracy. It is also similar to the $\mathbb{Z}_3$ symmetry in 3-state Potts model under Fradkin-Kadanoff transformation.
 
 Now we extend the discussion from $\mathbb{Z}_3$ to $\mathbb{Z}_p$ case. The Yang-Baxter operator associated with $\mathbb{Z}_p$ parafermion has 
been proposed for years~\cite{fateev1982self,jones1989on}. Also, the associated $p$-level and $N$-partite entangled states have been well studied, both 
from the parameter-independent~\cite{wang2014multipartite} and parameter-dependent Yang-Baxter 
operators~\cite{wojcik2003generation,rowell2010extraspecial,rowell2016parameter}. In the recent work~\cite{rowell2016parameter}, the author constructed 
the parametrized Gaussian N-partite generalized Yang-Baxter operators, and discussed the associated N-partite entangled qupit states, which is a 
generalized work of reference~\cite{rowell2010extraspecial}. Now we analyse the $\mathbb{Z}_p$ case based on the reference~\cite{rowell2016parameter}. For the 2-partite case, acting the parameter $a$ in Yang-Baxter operator $\tilde{R}(a)$(see reference~\cite{rowell2016parameter}, equation 2.6) on the 2-qupit natural basis, one 
obtains some states similar to the following form,
\begin{equation}
\tilde{R}(a)|00\rangle=\frac{1}{\sqrt{p}}\sum_{j=0}^{p-1}\tilde{X}_j(a)|j\rangle\otimes|j\rangle,
\end{equation}  
where $\tilde{X}_j(a)$ is a function of parameter $a$ and comes from the reference~\cite{rowell2016parameter}. When $a=0$, the $\tilde{R}(a)$-matrix reduces 
to the braid matrix, $|\tilde{X}_j(a)|=1$,  and it is easy to check that the associated 2-quipt states achieve their maximal entangled degree, where the 
entangled degree is defined by von Neumann entropy or the $\ell_1$-norm defined in our previous sections. For the $N$-partite case, when $a=0$, the 
similar $|\tilde{X}_j(0)=1|$ for all $j$ occurs and we conclude that the $N$-qupit entangled states also coincide with the related physical quantities we 
defined above. Indeed, both the 2-partite and  the $N$-partite case are determined by the parameter $\tilde{X}_j(a)$, hence it is reasonable that their 
entangled properties are similar.

But we should emphasize that for $p>4$, the quantum dimension $d>2$, hence the 2-partite $\breve{R}$-matrix obtained from our  Yang-Baxterization  is no longer unitary.

\section{Yang-Baxter Equation with Different Spins}\label{sec4}
%
%
 In this section, we mainly focus on the Yang-Baxter equation with different spins. 
It is well known that the YBE describes the 3-particle scattering processing, and the three particles are described in three tensor lattice spaces.  Originally, the YBE does not limit the three particles to have the same lattice spaces. It is just for simplicity that people usually adopt  three same lattice spaces for YBE. Indeed, the corresponding braid group representations can be constructed by quantum groups through the standard theory of Drinfeld and Reshetikhin~\cite{drinfeld1986quantum,drinfeld1985dokl,reshetikhin1988quantized,kirillow1989representations}. For the
 weight conservation solutions to YBE in different lattice spaces, series of results have been made~\cite{ge1991general,ge1992yang}.
Inspired by the Bell matrix in describing quantum entanglement, here we will discuss the solution to the YBE in different lattice spaces without weight conservation. 

Let us start with the ``braid-type'' relation: three strands with spin indices $\{1,\frac{1}{2},\frac{1}{2}\}$. For consistence, the ``braid-type'' relation are constrained by three relations.
\begin{eqnarray*}
(1). &S^{1\frac{1}{2}}_{12}S^{1\frac{1}{2}}_{23}S^{\frac{1}{2}\frac{1}{2}}_{12}=S^{\frac{1}{2}\frac{1}{2}}_{23}S^{1\frac{1}{2}}_{12}S^{1\frac{1}{2}}_{23};\\
(2). &S^{\frac{1}{2}1}_{12}S^{\frac{1}{2}\frac{1}{2}}_{23}S^{1\frac{1}{2}}_{12}=S^{1\frac{1}{2}}_{23}S^{\frac{1}{2}\frac{1}{2}}_{12}S^{\frac{1}{2}1}_{23};\\
(3). &S^{\frac{1}{2}\frac{1}{2}}_{12}S^{\frac{1}{2}1}_{23}S^{\frac{1}{2}1}_{12}=S^{\frac{1}{2}1}_{23}S^{\frac{1}{2}1}_{12}S^{\frac{1}{2}\frac{1}{2}}_{23}.
\end{eqnarray*} 
Whose space dimension are spanned in : $3\times 2\times2$. Relations 1-3 can be easily understood in the graphical forms:
\begin{equation}
(1).\begin{tikzpicture}[baseline=-\dimexpr\fontdimen22\textfont2\relax]
\draw(0,-1.1)node{\scriptsize $1$}(0.36,-1.1)node{\scriptsize $\frac{1}{2}$}(0.71,-1.1)node{\scriptsize $\frac{1}{2}$};
\braid[line width=1.3pt, height=15pt, width=10pt, number of strands=3, rotate=0,gray, style strands={3}{line width=1.7pt,red}] (braid) at (0,1.3) s_1^{-1}s_2^{-1}s_1^{-1}; 
\end{tikzpicture}=
\begin{tikzpicture}[baseline=-\dimexpr\fontdimen22\textfont2\relax]
\draw(0,-1.1)node{\scriptsize $1$}(0.36,-1.1)node{\scriptsize $\frac{1}{2}$}(0.71,-1.1)node{\scriptsize $\frac{1}{2}$};
\braid[line width=1.3pt, height=15pt, width=10pt, number of strands=3, rotate=0,gray, style strands={3}{line width=1.7pt,red}] (braid) at (0,1.3) s_2^{-1}s_1^{-1}s_2^{-1}; 
\end{tikzpicture};
(2).\begin{tikzpicture}[baseline=-\dimexpr\fontdimen22\textfont2\relax]
\draw(0,-1.1)node{\scriptsize $\frac{1}{2}$ }(0.36,-1.1)node{\scriptsize $1$}(0.71,-1.1)node{\scriptsize $\frac{1}{2}$};
\braid[line width=1.3pt, height=15pt, width=10pt, number of strands=3, rotate=0,gray, style strands={2}{line width=1.7pt,red}] (braid) at (0,1.3) s_1^{-1}s_2^{-1}s_1^{-1}; 
\end{tikzpicture}=
\begin{tikzpicture}[baseline=-\dimexpr\fontdimen22\textfont2\relax]
\draw(0,-1.1)node{\scriptsize $\frac{1}{2}$}(0.36,-1.1)node{\scriptsize $1$ }(0.71,-1.1)node{\scriptsize $\frac{1}{2}$};
\braid[line width=1.3pt, height=15pt, width=10pt, number of strands=3, rotate=0,gray, style strands={2}{line width=1.7pt,red}] (braid) at (0,1.3) s_2^{-1}s_1^{-1}s_2^{-1}; 
\end{tikzpicture};
(3).\begin{tikzpicture}[baseline=-\dimexpr\fontdimen22\textfont2\relax]
\draw(0,-1.1)node{\scriptsize $\frac{1}{2}$}(0.36,-1.1)node{\scriptsize $\frac{1}{2}$}(0.71,-1.1)node{\scriptsize $1$ };
\braid[line width=1.3pt, height=15pt, width=10pt, number of strands=3, rotate=0,gray, style strands={1}{line width=1.7pt,red}] (braid) at (0,1.3) s_1^{-1}s_2^{-1}s_1^{-1}; 
\end{tikzpicture}=
\begin{tikzpicture}[baseline=-\dimexpr\fontdimen22\textfont2\relax]
\draw(0,-1.1)node{\scriptsize $\frac{1}{2}$}(0.36,-1.1)node{\scriptsize $\frac{1}{2}$}(0.71,-1.1)node{\scriptsize $1$};
\braid[line width=1.3pt, height=15pt, width=10pt, number of strands=3, rotate=0,gray, style strands={1}{line width=1.7pt,red}] (braid) at (0,1.3) s_2^{-1}s_1^{-1}s_2^{-1}; 
\end{tikzpicture}.
\end{equation}

One type of local unitary solutions to the relations 1-3 is expressed as follows
\begin{equation}
\begin{aligned}
&S^{1\frac{1}{2}}_{12}=B^{(6)}\otimes I^{(2)},\quad S^{1\frac{1}{2}}_{23}=I^{(2)}\otimes B^{(6)},\quad S^{\frac{1}{2}\frac{1}{2}}_{12}=B^{(4)}\otimes I^{(3)},\\
& S^{\frac{1}{2}\frac{1}{2}}_{23}=I^{(3)}\otimes B^{(4)}, \quad
 S^{\frac{1}{2}1}_{12}=[B^{(6)}]^{OT}\otimes I^{(2)},\quad S^{\frac{1}{2}1}_{23}=I^{(2)}\otimes [B^{(6)}]^{OT}.
\end{aligned}
\end{equation}
Where $B^{(4)}$ is $4\times4$ Bell braid matrix  in spin-$\{\frac{1}{2},\frac{1}{2}\}$ space,  $B^{(6)}$ is $6\times6$ braid matrix in spin-$\{1, \frac{1}{2}\}$ space, $[B^{(6)}]^{OT}$ is the off-diagonal transpose of $B^{(6)}$ and is represented in spin-$\{\frac{1}{2},1\}$ space, $I^{(2)}$ and $I^{(3)}$ are identity matrices. The concrete forms of braid matrices are expressed as follows,
\begin{equation}
\begin{aligned}
&B^{(6)}=\frac{1}{\sqrt{2}}\left[\begin{matrix}
1 & 0 & 0 & 0 & 0 & e^{i\phi} \\
0 & 0 & 1 & 1 & 0 & 0\\
0 & \sqrt{2} & 0 & 0 & 0 & 0\\
0 & 0 & 0 & 0 & \sqrt{2} & 0\\
0 & 0 & -1 & 1 & 0 & 0\\
-e^{-i\phi}  & 0 & 0 & 0 & 0 & 1
\end{matrix}\right], \\
&B^{(4)}=\frac{1}{\sqrt{2}}\left[\begin{matrix}
1 & 0 & 0 & e^{i\phi} \\
0 & 1 & 1 & 0\\
0 & -1 & 1 & 0 \\
-e^{-i\phi}  & 0 & 0 & 1 
\end{matrix}\right]. \textrm{(Bell matrix)}
\end{aligned}
\end{equation}

Yang-Baxterization of the ``braid-type'' relation leads to  the solutions $\breve{R}^{(4)}$ and  $\breve{R}^{(6)}$,
\begin{align}
&&\breve{R}^{1\frac{1}{2}}_{12}(\theta_1)\breve{R}^{1\frac{1}{2}}_{23}(\theta_2)\breve{R}^{\frac{1}{2}\frac{1}{2}}_{12}(\theta_3)=\breve{R}^{\frac{1}{2}\frac{1}{2}}_{23}(\theta_3)\breve{R}^{1\frac{1}{2}}_{12}(\theta_2)\breve{R}^{1\frac{1}{2}}_{23}(\theta_1),\\
&&\breve{R}^{\frac{1}{2}1}_{12}(\theta_1)\breve{R}^{\frac{1}{2}\frac{1}{2}}_{23}(\theta_2)\breve{R}^{1\frac{1}{2}}_{12}(\theta_3)=\breve{R}^{1\frac{1}{2}}_{23}(\theta_3)\breve{R}^{\frac{1}{2}\frac{1}{2}}_{12}(\theta_2)\breve{R}^{\frac{1}{2}1}_{23}(\theta_1),\\
&&\breve{R}^{\frac{1}{2}\frac{1}{2}}_{12}(\theta_1)\breve{R}^{\frac{1}{2}1}_{23}(\theta_2)\breve{R}^{\frac{1}{2}1}_{12}(\theta_3)=\breve{R}^{\frac{1}{2}1}_{23}(\theta_3)\breve{R}^{\frac{1}{2}1}_{12}(\theta_2)\breve{R}^{\frac{1}{2}\frac{1}{2}}_{23}(\theta_1).
\end{align}
where
\begin{equation}
\begin{aligned}
&\breve{R}^{1\frac{1}{2}}_{12}(\theta)=\breve{R}^{(6)}(\theta)\otimes I^{(2)},\quad \breve{R}^{1\frac{1}{2}}_{23}=I^{(2)}\otimes \breve{R}^{(6)}(\theta),\\
& \breve{R}^{\frac{1}{2}\frac{1}{2}}_{12}=\breve{R}^{(4)}(\theta)\otimes I^{(3)},\quad \breve{R}^{\frac{1}{2}\frac{1}{2}}_{23}(\theta)=I^{(3)}\otimes \breve{R}^{(4)}(\theta),\\
&\breve{R}^{\frac{1}{2}1}_{12}(\theta)=[\breve{R}^{(6)}(\theta)]^{OT}\otimes I^{(2)},\quad \breve{R}^{\frac{1}{2}1}_{23}=I^{(2)}\otimes [\breve{R}^{(6)}(\theta)]^{OT}.
\end{aligned}
\end{equation}
The concrete forms of the Yang-Baxter solutions are
 \begin{eqnarray*}
&&\breve{R}^{(6)}(\theta)=\left[\begin{matrix}
\cos\theta & 0 & 0 & 0 & 0 & \sin\theta e^{i\phi}\\
0 & 0 & \cos\theta &  \sin\theta & 0 & 0\\
0 & 1 & 0 & 0 & 0 & 0\\
0 & 0 & 0 & 0 & 1 & 0\\
0 & 0 &  -\sin\theta & \cos\theta & 0 & 0\\
-\sin\theta e^{-i\phi} & 0 & 0 & 0 & 0 & \cos\theta
\end{matrix}\right], \\ 
&& \breve{R}^{(4)}(\theta)=\left[\begin{matrix}
\cos\theta & 0 & 0 & \sin\theta e^{i\phi} \\
0 & \cos\theta & \sin\theta & 0\\
0 & -\sin\theta & \cos\theta & 0 \\
-\sin\theta e^{-i\phi}  & 0 & 0 & \cos\theta 
\end{matrix}\right], \quad \text{(Type-II)}.
\end{eqnarray*}
Here we note that the $4\times4$ $\breve{R}$-matrix is still the type-II solution to our previous standard YBE with the same spin spaces. Both $4\times4$ $\breve{R}$ and $6\times6$ $\breve{R}$ do not hold weight conservation. Correspondingly, the parameter constraint for YBE is: 
\begin{equation*}
\tan\theta_2=\frac{\tan\theta_1+\tan\theta_3}{1+\tan\theta_1\tan\theta_3}.
\end{equation*}
 This is still the Lorentzian additivity.  When $\theta_i$'s$=\frac{\pi}{4}$,  the $\breve{R}(\theta)$-matrices turn out to be the braid matrices. 
 
 The role of $4\times4$ $\breve{R}$-matrix in quantum entanglement has been investigated. Now let us briefly introduce the role of $6\times6$ $\breve{R}$-matrix in generating qubit-qutrit quantum entanglement.
Applying $\breve{R}^{1\frac{1}{2}}(\theta)=\breve{R}^{(6)}(\theta)\otimes I^{(2)}$ on qubit-qutrit natural bases $\{|\tfrac{1}{2}\rangle_1|1\rangle_2,|\tfrac{1}{2}\rangle_1|0\rangle_2,|\tfrac{1}{2}\rangle_1|-1\rangle_2,|-\tfrac{1}{2}\rangle_1|1\rangle_2,|-\tfrac{1}{2}\rangle_1|0\rangle_2,|-\tfrac{1}{2}\rangle_1|-1\rangle_2\}$, one obtains
 \begin{align}
&\breve{R}_{12}^{1\frac{1}{2}}(\theta)|\tfrac{1}{2}\rangle_1|1\rangle_2=\cos\theta|1\rangle_1|\tfrac{1}{2}\rangle_2-\sin\theta e^{-i\phi}|-1\rangle_1|-\tfrac{1}{2}\rangle_2,\\
&\breve{R}_{12}^{1\frac{1}{2}}(\theta)|-\tfrac{1}{2}\rangle_1|-1\rangle_2=\cos\theta|-1\rangle_1|-\tfrac{1}{2}\rangle_2+\sin\theta e^{i\phi}|1\rangle_1|\tfrac{1}{2}\rangle_2,\\
&\breve{R}_{12}^{1\frac{1}{2}}(\theta)|\tfrac{1}{2}\rangle_1|-1\rangle_2=\cos\theta|1\rangle_1|-\tfrac{1}{2}\rangle_2-\sin\theta |-1\rangle_1|\tfrac{1}{2}\rangle_2,\\
&\breve{R}_{12}^{1\frac{1}{2}}(\theta)|-\tfrac{1}{2}\rangle_1|1\rangle_2=\cos\theta|1\rangle_1|-\tfrac{1}{2}\rangle_2+\sin\theta |-1\rangle_1|\tfrac{1}{2}\rangle_2,\\
&\breve{R}_{12}^{1\frac{1}{2}}(\theta)|\tfrac{1}{2}\rangle_1|0\rangle_2= |0\rangle_1 |\tfrac{1}{2}\rangle_2,\\
&\breve{R}_{12}^{1\frac{1}{2}}(\theta)|-\tfrac{1}{2}\rangle_1|0\rangle_2= |0\rangle_1 |-\tfrac{1}{2}\rangle_2.
\end{align}
We find that $\breve{R}^{(6)}$ entangles the qubit states $|\pm\frac{1}{2}\rangle$ and qutrit states $|\pm1\rangle$, while the qutrit state $|0\rangle$ is not entangled. Hence the physical meaning of $\theta$ in $\breve{R}^{(6)}$ is continuous qubit-qutrit entangled degree. When $\theta=\pi/4$, the entangled states have the maximal entangled degree.
 
 Besides generating qubit-qutrit entanglement, the $6\times6$ $\breve{R}$-matrix  also has much richer physics in potential, such as in generating many body models for mixed qubit-qutrit systems~\cite{drillon1983classical}, in describing spin-orbital optical vortex beam~\cite{bliokh2015spin}, and so on.

\section{Conclusion and Discussion}
In summary, we obtain the new type of solutions to YBE, both in standard case(three same spin lattices) and non-standard case(different spin lattices). These new types of solutions come from the Yang-Baxterization of braid matrices, and the parameters obey Lorentzian-like additivity. Based on such solutions, the new roles of YBE in quantum information, anyon system, and $\ell_1$-norm are discussed. These new results  can never be obtained from the type-I solutions of YBE, which obey the Galilean additivity. 


For better understanding about the new types of solutions, let us make a comparison between the type-I solutions and the type-II solutions, as shown in table~\ref{tab1}.
\begin{table}[h]
\caption{\label{ex}A comparison between the  type-I solutions and the type-II solutions.}\label{tab1}
\begin{center}
\begin{tabular}{l|l}
\br
 Type-I &  Type-II\\
\mr
Lie algebra, say $SO(N)$, $SU(N)$... & Extra special-2 group($4\times4$), or $SU(N)$...\\
\mr
Galilean additivity $u_2=u_1+u_3$. & Lorentzian additivity $u_2=\frac{u_1+u_3}{1+u_1u_3}$. \\
\mr
Exact solvable models. & Quantum entanglement. \\
\mr
6-vertex, Heisenberg chain: $S^+_iS^-_{i+1}$. & 8-vertex, superconductor: $a^{+}_i a^+_{i+1}$. \\  
\mr
RTT:  Yangian (Rational $\breve{R}$). & Most possibly no RTT relation.\\
\mr
{Temperley-Lieb algebra $d=2$}. & Temperley-Lieb algebra $d=\sqrt{2}$ or $\sqrt{3}$.\\
\mr
Min $\ell_1$-norm for $D^{\frac{1}{2}}$-matrix. & Max $\ell_1$-norm for $D^{\frac{1}{2}}$-matrix.\\
\mr
Discrete symmetry $C,\,P,\,T$. &{Besides  $C,\,P,\,T$,  collective discrete symmetry $\Gamma^{2(3)}=1$}.\\
\br
\end{tabular}
\end{center}
\end{table}

At last, we would like to put forward two interesting problems. The first one is how to understand the Lorentzian additivity for type-II solution. One possible candidate is in describing the velocity of anyon. For example, for the $4\times4$ type-II, it has the equivalent Majorana representation, which corresponds to the Ising anyon system. By now, physicists suppose that all  of the anyons possess light velocity in the vacuum. How to extend the definition of anyon velocities and  make connections between the anyon velocities and YBE are still challenging. The second problem is about the non-standard YBE and its applications, such as in describing scattering processing of different types of particles, or in describing the interactions between different physical quantities. Overall, it is still challenging in pursuing new physical meanings of YBE.

\ack{This work is in part supported by NSF of China (Grant No. 11475088).}
\section*{References}

\end{document}